\documentclass[aps,apl,floatfix,twocolumn,tightenlines,amsmath,amssymb]{revtex4}

\usepackage{graphicx}

\usepackage{amsmath}

\usepackage{amssymb}
\newcommand*{\rom}[1]{\expandafter\@slowromancap\romannumeral #1@}

\usepackage{epstopdf}

\usepackage{color}

\newcommand{\ketbra}[2] {{|#1 \rangle\!\langle #2|}}

\begin{document}

\title{Mapping the Chemical Potential Landscape of a Triple Quantum Dot}
\author{M. A. Broome, S. K. Gorman, J. G. Keizer, T. F. Watson, S. J. Hile, W. J. Baker, M. Y. Simmons}
\affiliation{Centre of Excellence for Quantum Computation and Communication Technology, School of Physics, University of New South Wales, Sydney, New South Wales 2052, Australia}
\date{\today}

\begin{abstract}

We investigate the non-equilibrium charge dynamics of a triple quantum dot and demonstrate how electron transport through these systems can give rise to non-trivial tunnelling paths. Using a real-time charge sensing method we establish tunnelling pathways taken by particular electrons under well-defined electrostatic configurations. We show how these measurements map to the chemical potentials for different charge states across the system. We use a modified Hubbard Hamiltonian to describe the system dynamics and show that it reproduces all experimental observations. 

\end{abstract}

\maketitle
\section{Introduction}
The sensing of individual electron charges confined to quantum dots (QD) forms the basis of many important quantum mechanical measurements in solid state systems such as spin readout, transfer and coherent manipulation~\cite{dicarlo2004,braakman2013,cao2012,watson2015}. As the complexity of coupled QD devices increases so does the electronic tunnelling processes that occur within them~\cite{schleser2004,PhysRevLett.112.176803,watson2015b}. In particular, for multi-QD systems the tunnelling of electrons from dot-to-dot will often occur in multiple stages resulting in non-trivial tunnelling paths. Establishing the path an individual electron takes is important when quantum information is being stored on specific particles within a multi-QD system~\cite{morello2010,buch2013,A.:2016fk}.

In any multi-QD architecture, measurement of the steady state charge occupation via a charge stability map~\cite{PhysRevB.67.161308,petta2004manipulation,PhysRevLett.94.196802,A.:2016fk,taylor2007} is a prerequisite for more complex measurements such as spin readout~\cite{elzerman2004,morello2010,buch2013}. However, these maps only present the long-term charge equilibrium of the system. This is because a charge stability measurement reveals information about the chemical potential of quantum dots with respect to their electron reservoir, not with respect to one another leaving the tunnelling path taken by an individual electron through the system unknown, see Fig.~\ref{fig:device}~\cite{hanson2007}. In order to determine the actual tunnelling path a particular electron has taken it is necessary to perform time resolved charge sensing~\cite{kung2009} thereby measuring the non-equilibrium dynamics of the QD system~\cite{house2011}. 

The smallest system where non-trivial electron tunnelling paths can occur is a triple quantum dot (TQD). Since their first realisation, a host of new physics has been investigated using the TQD~\cite{PhysRevLett.75.705} including new tunnelling regimes~\cite{watson2014,PhysRevB.73.235310,PhysRevLett.112.176803}, novel spin-blockade effects~\cite{busl2013}, and complex coherent spin physics~\cite{gaudreau2012,laird2010,A.:2016fk} for use in quantum information. In this paper we investigate the non-equilibrium behaviour of single-electron tunnelling in a TQD fabricated using precision placed donor atoms in silicon~\cite{fuechsle2012,watson2014,gorman2015b}. We use real-time charge sensing to determine the preferred tunnelling paths of electrons over a range of electrostatic potentials. Importantly we develop a Anderson-Hubbard model that describes the various electron pathways through multi-QD structures.

The paper is set out as follows: In section~\ref{sec:the} we outline the model used to describe the electronic tunnelling pathways in multi-QD systems. Importantly, the device has been designed so that dot-dot tunnel rates are much greater than dot-reservoir tunnel rates. This means that the individual dot-reservoir rates are the limiting inputs for this model. From it we show we can accurately determine the tunnelling pathways under any electrostatic configuration. In section~\ref{sec:exp} we describe the device design and fabrication used to demonstrate our real-time charge sensing technique, the results of which are presented in section~\ref{sec:res}. Finally in section~\ref{sec:dis} we discuss our results in the context of an increasingly complex QD architecture and show how our results can be used to elucidate tunnelling pathways in multi-electron systems.

\section{Theory of Multi-Dot Tunnelling}
\label{sec:the}

The most natural way to describe the tunnelling dynamics of electrons from multiple QDs is to use a network of coupled `sites', where each site has both an intra- and inter-site energy $\epsilon$ and $U$ respectively. For $M$ QDs each hosting at most $n{=}1$ electrons coupled to a single reservoir the system is embodied by a modified Hubbard Hamiltonian given by,
\begin{equation}
H = \sum^{M}_j (\epsilon_j - \mu_j) \hat{n}_j + \sum_{j{\neq}k} \frac{U_{j,k}}{2} \hat{n}_j \hat{n}_k + \sum_{i} \epsilon_i\hat{n}_i,
\end{equation}
where $\epsilon_j$ is the detuning, $\mu_j$ is the chemical potential of QD $j$, the number operator for QD $j$ is $\hat{n}_j$ and $U_{j,k}$ is the inter-Coulomb repulsion between QDs $j$ and $k$. The last term represents the electronic states of the reservoir. In this work we consider QDs on a linear graph with nearest-neighbour coupling, however, the model can be extended to an arbitrarily complex graph of QDs and reservoirs. The detunings $\epsilon_j$ are determined by voltages applied to gates, $V_g$, surrounding the device,
\begin{equation}
\label{eq:lever} 
\epsilon_j = \sum_{g} \alpha_{j,g} V_g,
\end{equation}
where $\alpha_{j,g}$ is the conversion factor (or lever arm) from applied voltage to energy from gate, $g$ to QD $j$~\cite{vanderwiel2003}. The complete electrostatic configuration is therefore deduced by considering all QD detunings which we will label $\Lambda{=}\{\epsilon_1,... \epsilon_j,... \epsilon_M\}$. We omit the coherent coupling terms between the QDs and instead assume that tunnelling between them is completely incoherent. This assumption is valid over the energy scale investigated in this work ($\sim10$~meV) which is much larger than regions where coherent tunnelling can occur ($\sim10$~$\mu$eV) i.e. at inter-dot transitions.

\begin{figure}[t]
\begin{center}
\includegraphics[width=1\columnwidth]{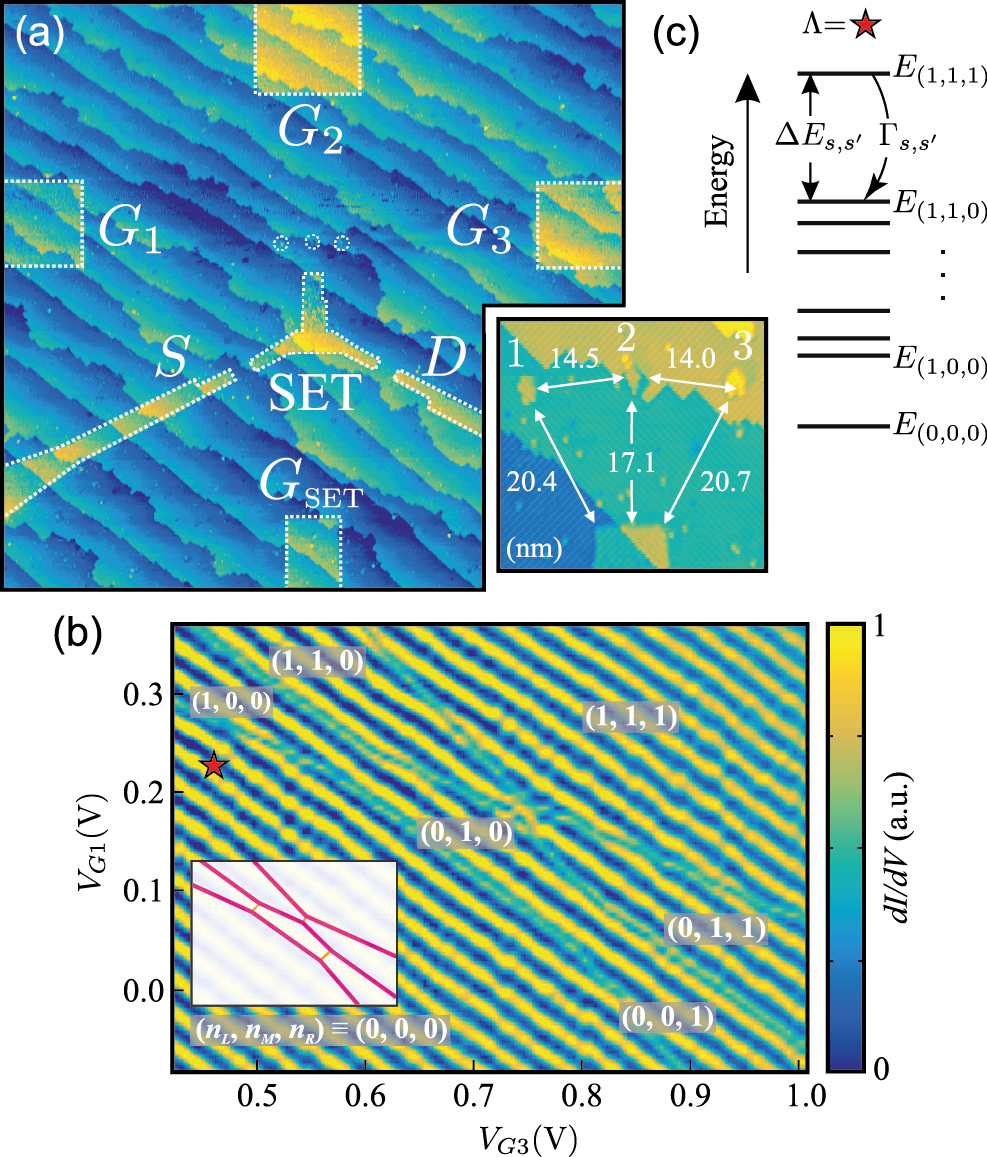}
\end{center}
\caption{{\bf A triple quantum dot in Si:P with an adjacent single-electron-transistor used as a charge sensor.} a) A scanning tunnelling micrograph of three small Si:P QDs incoherently tunnel coupled to a larger QD which itself is coupled to source ($S$) and drain ($D$) leads and acts as a sensing single electron transistor (SET). A gate $G_\text{SET}$ is used to control the SET while gates $\{G_{1},G_{2},G_{3}\}$ are used to control the QDs $\{1,2,3\}$. Inset shows a close up image of the three QDs. From the lithographic area it is estimated that they contain $\sim5$ donors each. The distances between the QDs as well as their individual distances to the SET are shown in nanometers. b) The conductance of the SET as a function of the gates $\{G_\text{1},G_\text{3}$\} for $G_\text{2}=0.9$~V and $G_\text{SET}=0.5$~V. Breaks in the Coulomb blockade peaks (lines running at $-45^\circ$) of the SET show where a charge transition from one of the QDs to the SET occurs. The inset provides a guide for where the charge transitions occur. The equivalent charge numbers on the QDs are given by $(n_1, n_2, n_3)$. c) A schematic representation of the modified Hubbard model showing the energies (not to scale) of different charge configurations at a detuning given by the star, also shown in (b). The incoherent tunnel rate $\Gamma_{s,s'}$ couples two charge configurations $s$ and $s'$. }
\label{fig:device}
\end{figure}

Finally, to incorporate the incoherent tunnelling in the system we transform the system from Hilbert space to Liouville space~\cite{gorman2015a} and introduce the incoherent rates $\Gamma_{s,s'}$ for the charge transition $s{\rightarrow}s'$. By doing this we can neglect the many charge states of the reservoir and consider only the $2^M$ possible charge states over the $M$ QDs. Importantly, the rates $\Gamma_{s,s'}$ represent direct transitions only, i.e. dot-to-dot or dot-to-reservoir, and do not represent any indirect tunnelling of an electron involving multiple tunnelling events. A schematic representation of the model is shown in Fig.~\ref{fig:device}c.

\begin{figure*}
\includegraphics[width=0.75\textwidth]{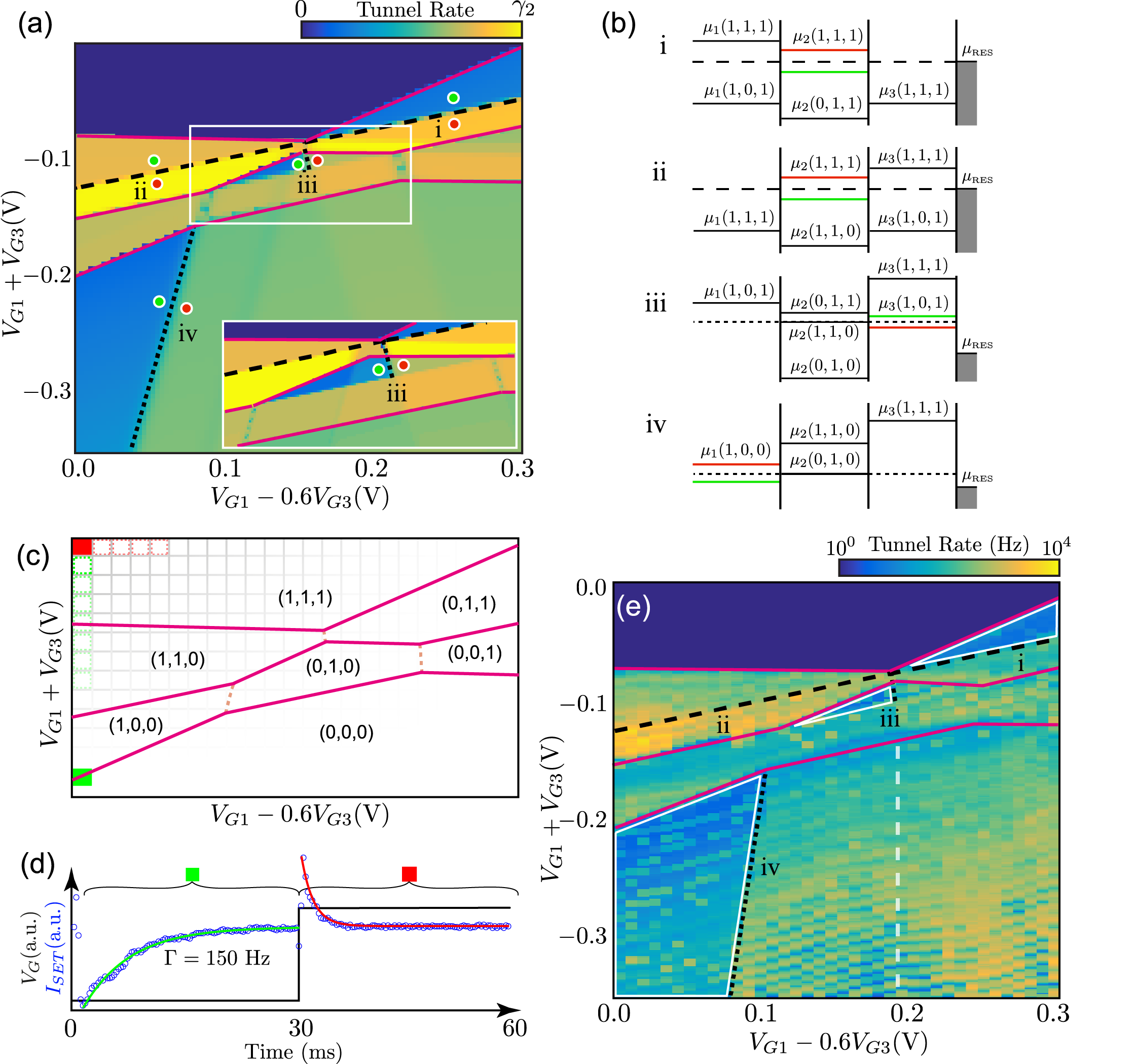}
\caption{{\bf Comparison of theoretical and experimental tunnel rate maps.} a) The theoretically predicted tunnel rate map for the TQD device scaled by the highest dot-reservoir tunnel rate, $\gamma_2$. Solid magenta lines show the charge transitions visible in standard charge stability maps (see Fig.~\ref{fig:device}b) while the dashed and dotted black lines correspond to transitions that can only be obtained from the tunnel rate map. The inset shows a zoom-in around the (0,1,0) charge region. b) The points (i), (ii), (iii), and (iv) corresponding to those shown in (a) are examined in terms of the chemical potentials of the QDs and the SET Fermi level (shown by the grey shaded region on the right of each panel). The levels shown in red and green correspond to the chemical potentials of those charge states at the points shown by the red and green circles in (a). The dashed lines in (i) and (ii) represent the SET Fermi level and corresponds to the dashed equivalent line shown in (a). Likewise, the dotted lines in (iii) and (iv) represent the relevant chemical potentials of charge states also shown in (a). c) The experimental procedure used to produce a tunnel rate map involves a two-level pulse from the (1,1,1) charge region (red square) to the detuning position $\Lambda$ (green square). The position marked by the red square is stepped across the $V_{G1}{-}0.6V_{G3}$ direction, while the final detuning position shown by the green grid is stepped across the charge transitions (magenta lines). d) The SET current is monitored in time to detect any electron tunnelling during the two level pulse. All data (blue circles) is comprised of 200 repetitions of the two level pulse (black line). A measure of the tunnel rate $\Gamma(\Lambda)$ is obtained from an exponential fit (green line) during the first phase of the two level pulse. The second phase of the pulse indicates the reloading of the electrons into the (1,1,1) charge region. e) A tunnel rate is obtained for different final detuning positions (green marker in (c)) $\Lambda$. There is a good qualitative agreement between the theoretical tunnel rate map shown in (a) which predicts all of the features obtained from the measurement. The areas enclosed by the white lines indicate the regions in gate space where an electron initially on QD-$1$ can only tunnel directly to the SET. The dashed white line indicates the cut used to calculate $\beta$ for Eq.~\ref{eq:rate}.}
\label{fig:theoryvsexpt}
\end{figure*}

These incoherent tunnel rates $\Gamma_{s,s'}$ follow a Fermi-distribution dependent on the difference in energy $\Delta E_{s,s'}$ between the charge states $s$ and $s'$,
\begin{equation}
\Gamma_{s,s'}(\Lambda){=}\begin{cases}
\frac{\eta_{j,k}}{1 + \textnormal{exp}[\Delta E_{s,s'}(\Lambda)/k_B T]}, & \Delta N = 0 \\ \\
\frac{\gamma_{j} - \beta(\Delta E_{s,s'}(\Lambda)-\mu_{\text{\tiny{RES}}})}{1 + \textnormal{exp}[(\Delta E_{s,s'}(\Lambda)-\mu_{\text{\tiny{RES}}})/k_B T]}, & \Delta N = 1,
\end{cases}
\label{eq:rate}
\end{equation}
\noindent where $N$ is the total number of electrons across all QDs; $\eta_{j,k}$ is the tunnel rate from QD $j{\rightarrow}k$ for a change $\Delta N{=}0$ and $\gamma_j$ is the tunnel rate from QD $j$ to the reservoir for $\Delta N{=}1$~\cite{gorman2015b}. Here we denote the chemical potential of the reservoir as $\mu_{\text{\tiny{RES}}}$. The thermal energy is $k_B T$ at a temperature, $T$, and the energy difference between two charge states, $s$ and $s'$, is given by, $\Delta E_{s,s'}(\Lambda){=}E_s(\Lambda){-}E_{s'}(\Lambda)$. To account for an observed linear increase in tunnel rates from the QDs to the reservoir as a function of detuning we include a phenomenological dimensionless factor, $\beta$, that increases the tunnel rate for large values of $\Delta E_{s,s'}(\Lambda){-}\mu_{\text{\tiny{RES}}}$. From a cut in the experimental data shown in Fig.~\ref{fig:theoryvsexpt}e as the white dashed line, we estimate this value to be, $\beta{=}2{\times}10^{-3}$. Following Fermi's golden rule, this factor accounts for the change in matrix element coupling the QD and the reservoir as the QD is detuned from the $\mu_j{=}\mu_{\text{\tiny{RES}}}$ condition~\cite{schleser2004}.

Using this model the tunnel rate from an initial excited charge state $\rho_i$ to the ground state $\rho_{gs}$ for the electrostatic configuration $\Lambda$ is determined by solving the Lindblad master equation~\cite{contreras-pulido2014,lindblad1976},
\begin{equation}
\frac{d\rho}{dt} = (\mathcal{L}_c + \mathcal{L}_{\Gamma})\rho,
\end{equation}
where $\mathcal{L}_c = i(\mathbb{I} \otimes H - H \otimes \mathbb{I})$ is the coherent time evolution and the incoherent term is given by, 
\begin{multline}
\mathcal{L}_{\Gamma} = \sum_{s,s'} \frac{\Gamma_{s,s'}(\Lambda)}{2} (2 L_{s,s'} \otimes L_{s,s'} - L_{s,s'}' L_{s,s'} \otimes \mathbb{I}\\ - \mathbb{I} \otimes L_{s,s'}' L_{s,s'}),
\end{multline}
where $L_{s,s'} = \ketbra{s}{s'}$. The thermal ground state of the system at $\Lambda$ is then determined by,
\begin{equation}
\rho_{gs}(\Lambda) = \frac{1}{Z} e^{-H(\Lambda)/k_B T},
\end{equation}
\noindent where, $\rho_{gs}(\Lambda)$ is the density operator for the thermal ground state and $Z{=}\textnormal{Tr}(e^{-H(\Lambda)/k_B T})$ is the partition function of the system. For our model we assume an electron temperature of $T{=}200$~mK, which is a typical value for our devices measured in a dilution fridge~\cite{buch2013,weber2014,gorman2015b}. After allowing the system to evolve from $\rho_i{\rightarrow}\rho_{gs}$, the tunnel time, $1/\Gamma(\Lambda)$ is taken to be the time when the probability of the charge ground state $\rho_{gs}$ reaches $1-e^{-1}$. By repeating this procedure under different electrostatic configurations, $\Lambda$, we can produce a so-called \textit{tunnel rate map} for the multi-QD system. Unlike the canonical charge stability map, the tunnel rate map can reveal information about the position of QD chemical potentials with respect to one another over the available gate range.

\section{Experiment}
\label{sec:exp}
The device studied in this work is a TQD fabricated using scanning tunnelling microscopy hydrogen lithography. The methods of this fabrication technique have been reported in detail previously~\cite{fuechsle2007,simmons2008,fuhrer2009}. The TQD device shown in Fig.~\ref{fig:device} is comprised of three small QDs, labelled $1$, $2$, and $3$ (left, middle, and right respectively), consisting of $\sim$\,5~P donors each, determined by examining the extent of the exposed lithographic area~\cite{buch2013,weber2014}. These small donor clusters are tunnel coupled to a large quantum dot made up of $\sim 1000$P atoms and placed at a distance of 17--21~nm from it, see Fig.~\ref{fig:device}a. In turn this larger dot is coupled to source ($S$) and drain ($D$) leads allowing electrons to flow across its tunnel junctions on either side, therefore acting as a single-electron-transistor (SET) charge sensor~\cite{kastner1992,fuechsle2012}. The SET  has a charging energy of $5{\pm}1$~meV and is operated with a source-drain bias of 0.3~mV. The electrostatics of the SET are controlled by the gate $G_{SET}$, whereas the QDs are predominantly tuned using the gates $G_1$, $G_2$ and $G_3$. The SET island serves as the electron reservoir in this system and the incoherent coupling rates of the QDs $\{1,2,3\}$ are given by \{$\gamma_{1},\gamma_{2},\gamma_{3}$\}, respectively.

For the experiments presented in this letter $G_2$ is used as a global gate to shift the potential of all the QDs, while $G_1$ and $G_3$ are used to detune the potential of the QDs with respect to the SET. As such, the relevant lever arms, $\alpha_{j,g}$ for Eq.~\ref{eq:lever} are those of $G_1$ and $G_3$. A charge stability diagram of the TQD system, showing the SET conductance as a function of $G_1$ and $G_3$, is presented in Fig.~\ref{fig:device}b. Lines running at ${\sim}45^{\circ}$ in the data show the Coulomb blockade of the SET and breaks in these lines correspond to charge transitions between the SET and the three QDs. Due to the different capacitive coupling of the gates $\{G_1,G_3\}$ to each of the QDs, three distinct lines of SET breaks with different slopes are visible in this gate space. In addition a characteristic pentagon structure associated with the quadruple point of a TQD around (0,1,0) can be seen (see inset of Fig.~\ref{fig:device})~\cite{busl2013,braakman2013,watson2014}. We note that the absolute electron number are not known for this device; however, for the purpose of this work we assign the charge states shown in Fig.~\ref{fig:device}b where ($n_1,n_2,n_3$) represent the electron numbers on QDs $1$, $2$ and $3$, respectively. This does not affect the physics discussed in this work.

\section{Results}
\label{sec:res}
Using the model described in section~\ref{sec:the} and initialising the system in the $s{=}(1,1,1)$ charge configuration we obtain a theoretical tunnel rate map for this device shown in Fig.~\ref{fig:theoryvsexpt}a. The electrostatic configuration $\Lambda$ in gate space $V_{G1}{-}0.6V_{G3}$ vs $V_{G1}{+}V_{G3}$ was determined by calculating the lever arms, $\alpha_{j,g}$ using a capacitance modelling program~\cite{weber2012a}. This theoretical tunnel rate map  takes as an input the individual tunnel rates of the three QDs to the SET reservoir which, due to both their different distances from the SET and donor numbers, are given approximately by $\gamma_{1}{=}150$~Hz, $\gamma_2{=}1200$~Hz and $\gamma_3{=}1000$~Hz (obtained from experiment). In another work on the same device~\cite{gorman2015b} the tunnel couplings between dots 1 and 2 and between dots 2 and 3 were measured to be $5.5$~GHz and $2.2$~GHz respectively, which are much greater than any dot-reservoir rate. Thus, as an input for the model it is sufficient to use a incoherent dot-dot rate that is much larger than any dot-reservoir rate in order to obtain the correct system dynamics, here we assume $\eta_{1,2}{=}\eta_{2,3}{=}1000\gamma_{1}$. This argument is valid for any device with inter-dot separations of $\sim$~15nm in donor based systems, which typically results in tunnel coupling rates of the order $10^6$-$10^9$~Hz~\cite{weber2014,House2015ng}. 

A tunnel rate map can be obtained experimentally following the procedure shown schematically in Fig.~\ref{fig:theoryvsexpt}c. At $t{=}0$ the device is initialised in the equivalent $(1,1,1)$ charge region, at the position labelled by the solid red square, from here an initial pulse is applied simultaneously to gates $G_1$ and $G_3$ to a position marked by the solid green square. During the first pulse phase, in cases where the ground state of the system contains a total electron number $<$3, one or more electrons will tunnel off the QDs to the SET reservoir during this pulse. Importantly, the length of this pulse, $t_1{=}30$~ms, is longer than any dot-to-reservoir tunnelling time, thus always allowing the system to reach its ground state charge configuration. A subsequent pulse is applied at time $t{=}t_1$ of length $t_2{=}30$~ms to reinitialise the charge state of the QDs into the ${(1,1,1)}$ charge configuration. Again, $t_2$ is much longer than any of the reservoir-to-dot tunnelling time. The same two level pulse sequence is repeated $200$ times and the average current of the SET is recorded over the duty cycle of one complete pulse sequence from $t{=}0$ to $(t_1{+}t_2)$, see Fig.~\ref{fig:theoryvsexpt}d.
 
In cases where at least one electron has tunnelled to the SET during the first phase of the pulse cycle the SET current will show an exponential change from which the experimental tunnel rate $\Gamma_{E}(\Lambda)$ is extracted. Note that, although theoretically multiple exponents of decay should be observed, in practice tunnelling will be dominated by the slowest rate, we therefore fit to a single exponential decay. The position of the first pulse is stepped across the charge stability region in segments shown by the subsequent green squares in Fig~\ref{fig:theoryvsexpt}c. In doing so, a tunnel rate can be deduced for every detuning position, $\Lambda$. Our measured tunnel rate map shown in Fig~\ref{fig:theoryvsexpt}e agrees very well with the theoretical map in Fig.~\ref{fig:theoryvsexpt}a. As well as showing the direct dot-reservoir charge transitions apparent in the standard charge stability map in Fig.~\ref{fig:device}b, the tunnel rate map reproduces features that arise from indirect tunnelling pathways.

As we show below, because the tunnel rates of the individual QDs to the SET reservoir have the relationship $\gamma_1{\ll}\gamma_3{<}\gamma_2$, in some circumstances it will be favourable for electrons to tunnel via other QDs as a means to reach the ground state in the shortest time possible, rather than tunnel directly to the SET.  Importantly, whether or not these types of tunnelling paths are allowed depends on the relative positions of the QD chemical potentials, which can be calculated using the constant interaction model~\cite{vanderwiel2003} (see Fig.~\ref{fig:theoryvsexpt}b). The positions of these chemical potentials can therefore be deduced directly from the tunnel rate map and we now discuss in detail the four points labelled in Fig.~\ref{fig:theoryvsexpt}a as (i-iv) that result in non-trivial tunnelling pathways. 

At position (i) in Fig.~\ref{fig:theoryvsexpt}a the ground-state charge configuration is (0,1,1) and the chemical potentials of the QDs and SET at the position shown by the red circle are related by $\mu_1\text{\small{(1,1,1)}}{>}\mu_2\text{\small{(1,1,1)}}{>}\mu_{\text{\tiny{RES}}}{>}\mu_3\text{\small{(1,1,1)}}$ (where the subscript refers to an electron transition from this dot), see Fig.~\ref{fig:theoryvsexpt}b. Since the tunnel rates for QD-$1$ and QD-$2$ are related by $\gamma_1{<<}\gamma_2$ an electron is much more likely to tunnel to the SET from QD-$2$ first before the tunnelling of an electron from QD-$1$. The charge state will now be $(1,0,1)$, leaving the electron on QD-$1$ free to tunnel across to QD-$2$ because $\mu_1\text{\small{(1,0,1)}}{>}\mu_2\text{\small{(0,1,1)}}$. That is, the system will effectively make the following state transitions: $(1,1,1){\rightarrow}(1,0,1){\rightarrow}(0,1,1)$. The reduction in tunnel rates at the point indicated by the green circle indicates where $\mu_2\text{\small{(1,1,1)}} < \mu_{\text{\tiny{RES}}}$ where a slower tunnel rate results because an electron can only tunnel to the SET from QD-$1$ at this point. The boundary between these two regions (black dashed line) indicates where $\mu_2\text{\small{(1,1,1)}}{=}\mu_{\text{\tiny{RES}}}$.

The situation at position (ii) is almost identical to that at position (i), only here the ground-state charge configuration is (1,1,0) i.e. QD-$3$ takes the place of QD-$1$ at position (i). Again, the interface between the regions at (ii) $\mu_2\text{\small{(1,1,1)}}{=}\mu_{\text{\tiny{RES}}}$ is given by the black dashed line. At the point labelled with the red circle the charge transitions $(1,1,1){\rightarrow}(1,0,1){\rightarrow}(1,1,0)$ are made, whereas at the green circle the electron initially on QD-$3$ tunnels directly to the SET. It is worth noting that the same $\mu_2\text{\small{(1,1,1)}}{=}\mu_{\text{\tiny{RES}}}$ line can be mapped all the way through from the (1,1,0) to the (0,1,1) charge region. 

At position (iii) the ground-state charge configuration is (0,1,0) and tunnelling from this region involves multiple tunnelling events over numerous possible pathways. Unlike at positions (i) and (ii) however, the tunnelling via different pathways depends on the relative \textit{inter-dot} chemical potentials. To illustrate this we will discuss one such tunnelling path in detail, the relevant energy diagram for which is shown in Fig.~\ref{fig:theoryvsexpt}b (iii). Here, $\mu_1\text{\small{(1,1,1)}}{>}\mu_2\text{\small{(1,1,1)}}{>}\mu_3\text{\small{(1,1,1)}}{>}\mu_{\text{\tiny{RES}}}$, i.e. any electron on any dot can in principle tunnel to the SET. In the case where QD-$3$ tunnels first, we are left with (1,1,0), and at the point shown by the red circle we have $\mu_2\text{\small{(1,1,0)}}{>}\mu_3\text{\small{(1,0,1)}}$ such that an electron will tunnel from QD-$2$ to QD-$3$. Finally, because $\mu_1\text{\small{(1,0,1)}}{>}\mu_2\text{\small{(0,1,1)}}$ the electron originally on QD-$1$ tunnels across to QD-$2$. The total electron movement in the system is: $(1,1,1){\rightarrow}(1,1,0){\rightarrow}(1,0,1){\rightarrow}(0,1,1){\rightarrow}(0,1,0)$. At the detuning position marked by the green circle however, we have $\mu_3\text{\small{(1,0,1)}}{>}\mu_2\text{\small{(1,1,0)}}$, which restricts the electron on QD-$2$ tunnelling over to QD-$3$ therefore forcing the electron on QD-$1$ to tunnel from there to the SET reservoir. This restriction reduces the observed tunnel rate because some of the tunnelling pathways will be limited by the slowest rate $\gamma_1$. A line separating two regions at (iii) indicates an alignment of the QD chemical potentials $\mu_2\text{\small{(1,1,0)}}{=}\mu_3\text{\small{(1,0,1)}}$.

The last position we discuss is far outside of the TQD quadruple point itself and is shown by position (iv) in Fig.~\ref{fig:theoryvsexpt}a. Here the ground state is (0,0,0), and between the green and red circles a transition can be seen where the two QD chemical potentials $\mu_1\text{\small{(1,0,0)}}{=}\mu_2\text{\small{(0,1,0)}}$, i.e. where an inter-dot electron transition can occur. Here, all three electrons can tunnel off to the SET, but the order in which they do so depends on the relative position of the QD chemical potentials with respect one another. At the position labeled by the green circle where $\mu_1\text{\small{(1,0,0)}}{<}\mu_2\text{\small{(0,1,0)}}$ the electron on QD-$2$ and $3$ tunnel to the SET first, followed by the slow tunnelling of the QD-$1$ electron at the rate $\gamma_1$. However, at the red circle after the electron from QD-$2$ has tunnelled the electron now residing on QD-$1$ can tunnel across to QD-$2$ where it will finally tunnel to the SET but now at the much faster rate of $\gamma_2$. A similar effect cannot be seen at the equivalent transition between QD-2 and QD-3 where $\mu_3\text{\small{(0,0,1)}}{=}\mu_2\text{\small{(0,1,0)}}$ because the tunnel rates, $\gamma_2$ and $\gamma_3$, are not different enough.

It is worth noting that the same information can be obtained from a tunnel rate map for an arbitrary set of tunnel rates, $\gamma_i$ despite different preferred tunnelling pathways. However, the visibility of the transitions will be reduced as the difference between the tunnel rates approaches zero.

\begin{figure}
\begin{center}
\includegraphics[width=1\columnwidth]{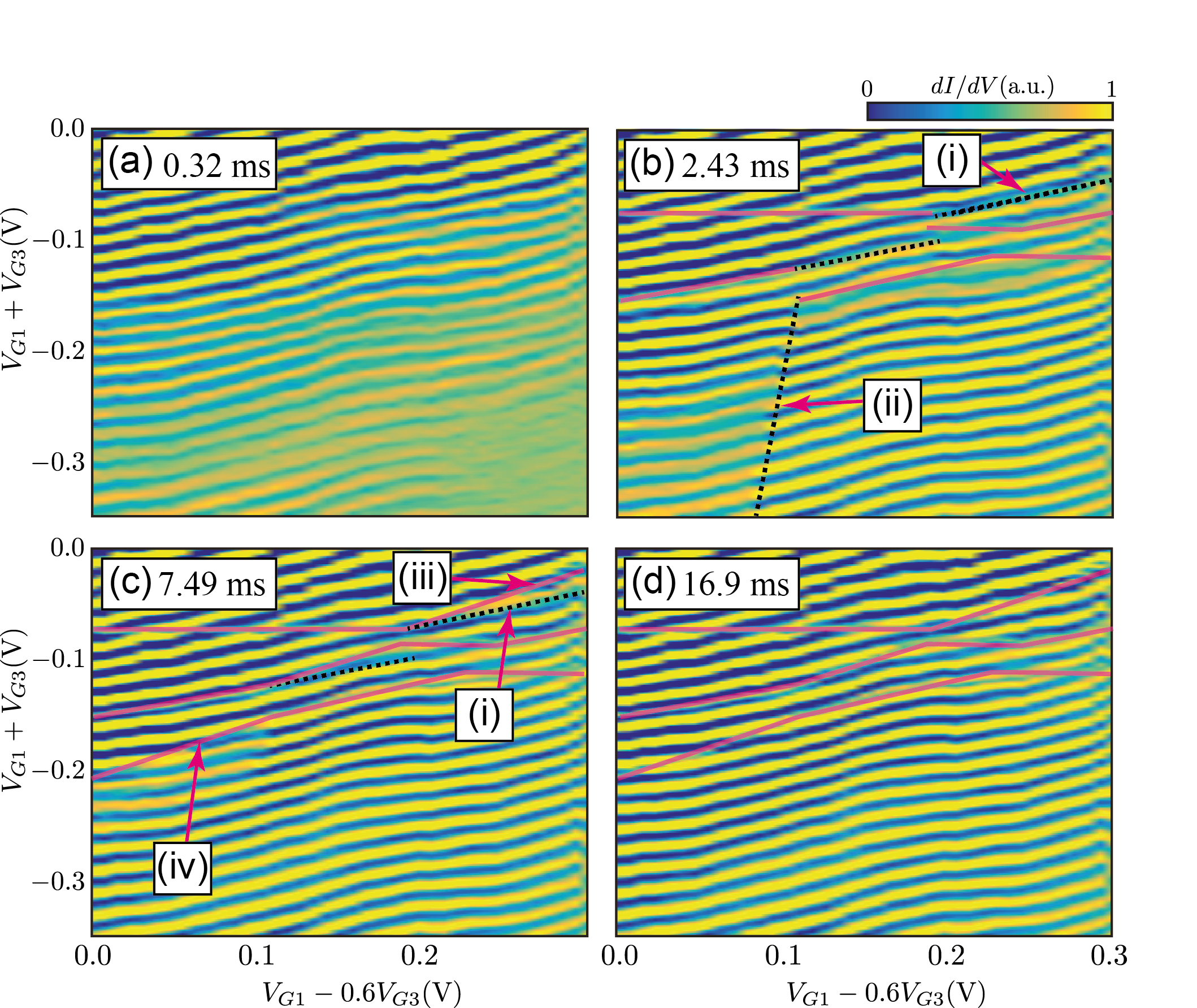}
\end{center}
\caption{{\bf Time-resolved charge stability maps for a triple quantum dot.} Maps showing the conductance of the SET charge sensor at different times from $t{=}0.32$~ms to $t{=}16.9$~ms. The solid magenta lines show those transitions that are apparent from a standard stability map, whereas the dashed lines are only visible using time resolved charge sensing. a) The conductance through the SET charge sensor at 0.32~ms before any electrons have tunnelled off the QDs. There are no SET breaks corresponding to QD transitions. b) At 2.43~ms, both the electrons from QD-$2$ and $3$ have tunnelled to the SET creating breaks in the Coulomb blockade where $\mu_2$ and $\mu_3$ are equal to $\mu_{\text{\tiny{RES}}}$. Point (i) shows where $\mu_2\text{\small{(1,1,1)}}{=}\mu_{\text{\tiny{RES}}}$; whereas point (ii) indicates the $\mu_1\text{\small{(1,0,0)}}{=}\mu_2\text{\small{(0,1,0)}}$ condition. c) At 7.49~ms, the electron on QD-$1$ begins to tunnel to the SET, thus two transitions in the upper right corner of the map are visible at points (iii) and (iv). At point (iii) the electron is tunnelling straight from QD-$1$ to the SET. The electron on QD-$1$ is also able to tunnel from the (1,0,0) charge state to the (0,0,0) on the bottom left of the map at point (iv). d) Finally, after a time 16.9~ms the map is equivalent to the standard charge stability map from Fig.~\ref{fig:device}b, as it shows all three of the QD transitions at equilibrium.}
\label{fig:gg_maps}
\end{figure}

To further reveal the non-equilibrium dynamics of electron movement across the TQD, we examine the instantaneous conductance through the SET charge sensor during the first pulse phase, between $t{=}0$ and $t_1$. These \textit{time resolved} charge stability maps are shown in Fig.~\ref{fig:gg_maps}. Figure~\ref{fig:gg_maps}a shows the instantaneous current through the SET at time $t{=}0.32$~ms. At this time no clear breaks in the SET Coulomb blockade peaks are apparent indicating that no electrons have tunnelled from the TQD system to the SET reservoir before this time, and therefore the system remains in the (1,1,1) charge configuration. In contrast, at $t{=}2.43$~ms, shown in panel (b), multiple SET breaks corresponding to charge transitions from the TQD to the SET reservoir can be seen. Not surprisingly, the line breaks observed at this time are associated with electron transitions from QD-$2$ and $3$ because they have the fastest tunnel rates.

Importantly, the map in Fig.~\ref{fig:gg_maps}b contains features seen in the standard charge stability map as well as some additional ones not seen in Fig.~\ref{fig:device}b that arise due to non-trivial tunnelling pathways. In particular an SET break corresponding to where the chemical potentials $\mu_2\text{\small{(1,1,1)}}{=}\mu_{\text{\tiny{RES}}}$ at point (i) in this map is clearly visible. This occurs since the electron on QD-$1$ has not had enough time to tunnel directly to the SET, yet the electron on QD-$2$ has tunnelled to the SET leaving space for the electron on QD-$1$ to tunnel across to this site (see discussion of (i) of Fig.~\ref{fig:theoryvsexpt}a above). Break (ii) in this map corresponds to where $\mu_1\text{\small{(1,0,0)}}{=}\mu_2\text{\small{(0,1,0)}}$ (see discussion of point (iv) from Fig.~\ref{fig:theoryvsexpt}a above). 

The map shown in Fig.~\ref{fig:gg_maps}c is taken at a time where the electron on QD-$1$ has started to tunnel to the SET reservoir, and as a result, in addition to break (i) we see another transition in the top right corner of this map, break (iii), where $\mu_1\text{\small{(1,1,1)}}{=}\mu_{\text{\tiny{RES}}}$.  Since we see both transitions at this time the system must have a non-zero probability of tunnelling via two pathways near this detuning region. In addition, another transition line corresponding to tunnelling from QD-$1$ is observed, at point (iv). Finally, after all the electrons have tunnelled at 16.9~ms in panel (d) the stability map of the TQD is fully recovered.

\section{Discussion}
\label{sec:dis}
In this work we consider the complex electron tunnelling pathways to a reservoir that arise in coupled QD systems. We have presented a detailed theoretical model that captures all of the non-equilibrium charge dynamics, and one that predicts very well the observed experimental signatures of non-trivial tunnelling processes. Although we consider only three QDs with nearest neighbour couplings, the model and results we present are applicable to any size and form of graph, including multiple QDs coupled to multiple reservoirs~\cite{seo2013,A.:2016fk,thalineau2012}. In these cases, one must add additional tunnel rates for individual reservoirs to Eq.~\ref{eq:rate}.  We also note that a reversal of our protocol, i.e. the \textit{loading} of electrons from a reservoir to the dots, will reveal an equivalent map of interdot chemical potentials. 

Our experimental method relies upon individual QDs having different tunnel rates to distinguish between different tunnelling pathways. For donor based architectures, this caveat is easily fulfilled due to the sensitive dependency of tunnel rates on dot-reservoir distances (sensitive to the order of the order of $1$~nm displacements). In gate defined QDs, incoherent tunnel rates can be tuned using external gate biases such that a sufficiently large difference between individual QDs may be attained~\cite{PhysRevLett.75.705}. However, any measurable difference in tunnel rate is sufficient in order to establish a tunnel rate map.

The time resolved charge stability maps presented in Fig.~\ref{fig:gg_maps} give a direct image of the inter-dot chemical potentials, which can later be used to elucidate the complex tunnelling pathways that occur in these systems. Importantly, electron pathways from dot-to-reservoir can be dependent on the inter-dot chemical potential structure, both deep inside a stable charge region as well as near dot-reservoir transitions. Understanding and controlling electron pathways is a vital component of multi-electron physics, and to this end tunnel rate maps provide an important characterisation tool for solid-state quantum information processing. For example, some spin readout schemes require the transfer of an electron from one QD to another before readout can be carried out~\cite{Veldhorst2015zt,A.:2016fk}. In these cases it becomes imperative to characterise the movement of specific electrons under all electrostatic conditions in order that the location and movement of quantum information in the system can tracked and processed correctly. 

\begin{acknowledgments}
This research was conducted by the Australian Research Council Centre of Excellence for Quantum Computation and Communication Technology (project no. CE110001027) and the US National Security Agency and US Army Research Office (contract no. W911NF-08-1-0527). M.Y.S. acknowledges an Australian Research Council Laureate Fellowship.
\end{acknowledgments}

\end{document}